\documentclass[journal=jpclcd,manuscript=letter]{achemso}
\usepackage{booktabs}
\usepackage{commath}
\usepackage{multirow}
\usepackage{dcolumn}
\usepackage{bm}
\usepackage{amsmath}
\usepackage{amssymb}
\usepackage{graphicx}
\usepackage[version=3]{mhchem}
\usepackage[T1]{fontenc}
\usepackage{subfigure}
\usepackage{array}
\usepackage{xr}
\usepackage{cases}
\usepackage{empheq}
\usepackage{longtable}
\usepackage[para,online,flushleft]{threeparttable}
\usepackage{xcolor}
\usepackage{placeins}
\usepackage{mhchem}

\makeatletter

\author{Jincheng Yu}
\affiliation{Department of Chemistry, Duke University, Durham, NC 27708, USA}
\author{Yuncai Mei}
\affiliation{Department of Chemistry, Duke University, Durham, NC 27708, USA}
\author{Zehua Chen}
\affiliation{Department of Chemistry, Duke University, Durham, NC 27708, USA}
\alsoaffiliation{Theoretical Chemistry Institute and Department of Chemistry, University of Wisconsin-Madison, Madison, WI 53706, USA}
\author{Weitao Yang}
\affiliation{Department of Chemistry, Duke University, Durham, NC 27708, USA}
\email{weitao.yang@duke.edu}
\alsoaffiliation{Department of Physics, Duke University, Durham, NC 27708, USA}

\title{
    Accurate Prediction of Core Level Binding Energies from Ground-State
    Density Functional Calculations: The Importance of Localization and 
    Screening
}

\makeatother

\begin{document}

\begin{tocentry}
\includegraphics[width=1\textwidth]{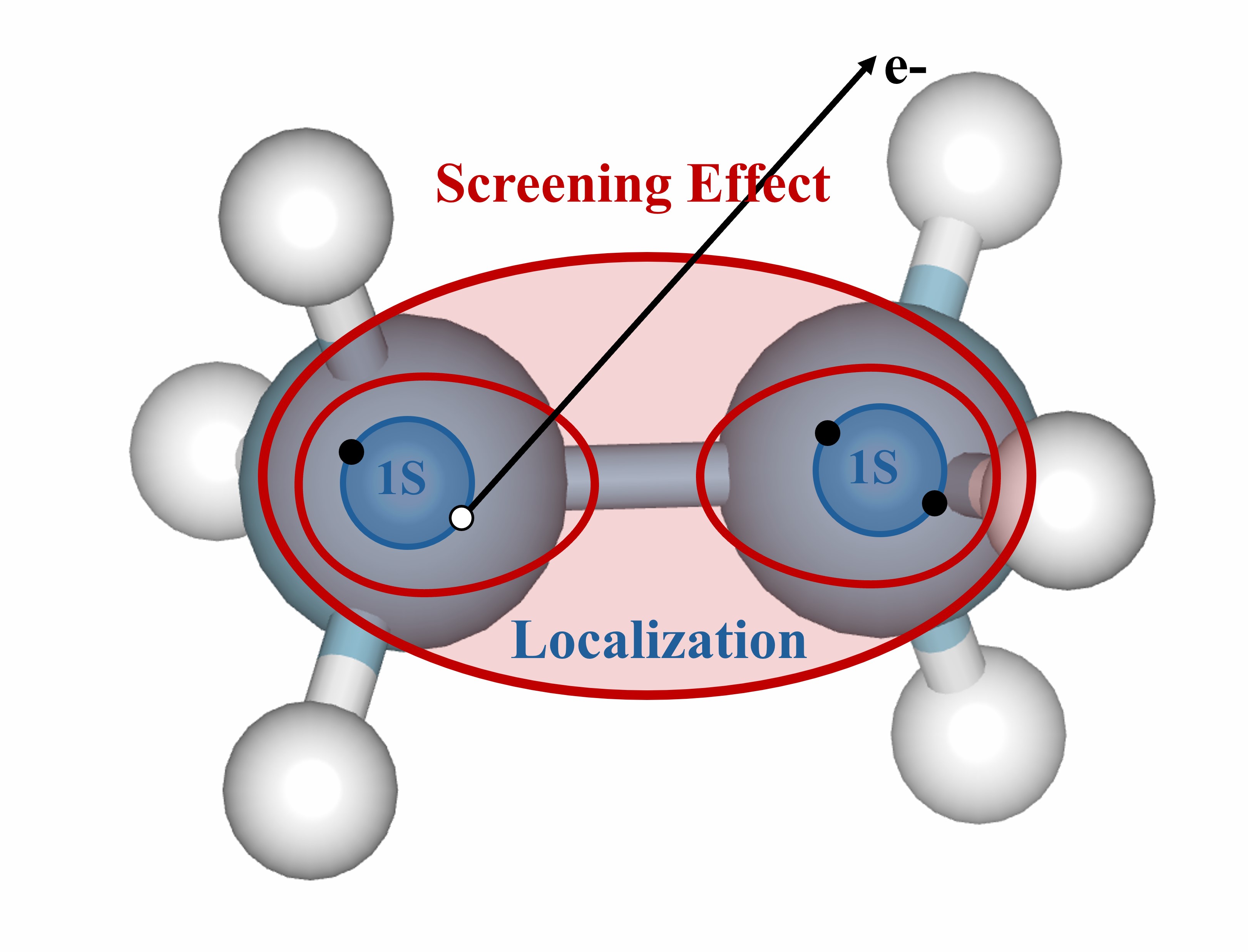}
\end{tocentry}

\begin{abstract}
A new method for predicting core level binding energies (CLBEs) is 
developed by both localizing the core-level states and describing the
screening effect.
CLBEs contain important information 
about the electronic structure, elemental chemistry, 
and chemical environment of molecules and materials. 
Theoretical study of CLBEs 
can provide insights for analyzing and interpreting the experimental results 
obtained from the X-ray photoelectron spectroscopy, 
in which the overlapping of signals is very common.
The localization of 
core-level holes is important for the theoretical calculation of CLBEs.
Predicting CLBEs from commonly used density functional approximations (DFAs) is challenging, 
because conventional DFAs often produce delocalized core-level states, 
especially when degenerate core-level states exist.
In this work,
we combine the localization procedure from the localized orbital 
scaling correction method and the curvature matrix generalized 
from the exact second-order correction method that contains the screening
effect, 
and the resulting approach
can accurately predict CLBEs from ground-state density functional 
calculations.
\end{abstract}

Core-level binding energies (CLBEs),
which are element-specific and sensitive to the chemical environment,
are crucial for understanding the
electronic properties, elemental chemistry, 
and oxidation state
of molecules and materials\cite{siegbahnElectronSpectroscopyAtoms1982,
bagusInterpretationXPSSpectra2013}.
Experimentally, 
CLBEs can be obtained from X-ray photoelectron 
spectroscopy (XPS)\cite{siegbahnVRaySpectroscopyPrecision1956,
siegbahn1967esca,barr1994principles}.
However, 
due to the overlap between peaks on the spectrum, 
assigning XPS signals precisely to the atomic sites can be challenging,
especially for complex materials\cite{aarvaUnderstandingXraySpectroscopy2019}.
Therefore, 
guidance from accurate theoretical analysis is often necessary for the
interpretation of experimental results\cite{
egelhoffCorelevelBindingenergyShifts1987,
bagusInterpretationXPSSpectra2013}.

Delta self-consistent field ($\Delta$SCF) 
method\cite{bagusSelfConsistentFieldWaveFunctions1965}
is one of the earliest developed approaches to compute CLBEs with high accuracy.
In $\Delta$SCF, CLBEs are calculated as the differences between the 
$N$-electron systems and the corresponding ($N-1$)-electron systems 
with a core-level hole. 
In general, 
relative CLBEs from $\Delta$SCF are close to experimental results, 
with the deviations being around 
0.2 eV for small systems\cite{pueyobellafontPerformanceMinnesotaFunctionals2016a}.
While the deviation has a small 
dependence on the exchange-correlation (XC) functional,
mean absolute errors (MAEs) from $\Delta$SCF are lower than 1 eV for most 
functionals\cite{
pueyobellafontPredictionCoreLevel2015,
kahkAccurateAbsoluteCoreelectron2019,
pueyobellafontPerformanceTPSSFunctional2016}.
Theoretically, 
calculating the binding energy by evaluating the total energies of 
the initial and final states remains challenging, 
particularly for bulk insulators\cite{ozakiAbsoluteBindingEnergies2017}. 
After the emission of a photoelectron, 
the system becomes non-periodic due to the formation of a core-level hole. 
This disruption of periodicity complicates the use of 
conventional electronic structure methods that rely on periodic boundary conditions. 
Additionally, 
the Coulomb potential of the ionized bulk cannot be addressed under periodic assumptions 
due to Coulombic divergence.
To address this issue,
several methods have been developed,
including neutralizing the final state by adding an electron into
the conduction band\cite{
pehlkeEvidenceSitesensitiveScreening1993,olovsson2010first,
ljungberg2011implementation,susi2015calculation},
approximating the bulk by a cluster model\cite{
bagus2013interpretation},
and using exact Coulomb cutoff method\cite{ozakiAbsoluteBindingEnergies2017}.
Numerical tests\cite{kahkCoreElectronBinding2018,
kahkAccurateAbsoluteCoreelectron2019,
pueyobellafontPredictingCoreLevel2017,
olovssonCorelevelShiftsFcc2005,kohlerDensityFunctionalStudy2004}
indicate the high accuracy of $\Delta$SCF.
Similar to $\Delta$SCF, 
other $\Delta$ approaches\cite{navesdebritoTheoreticalStudyRay1991,
agrenRelaxationCorrelationContributions1993,
trigueroSeparateStateVs1999,
suSecondOrderPerturbationTheory2016,
smigaSpinComponentScaledDMP2Parametrization2018,
nooijenDescriptionCoreExcitation1995,
ohtsukaInnershellIonizationsSatellites2006,
besleyEquationMotionCoupled2012,
senInclusionOrbitalRelaxation2018,
zhengPerformanceDeltaCoupledClusterMethods2019}
also require the explicit optimization of core-ionized or core-excited
($N-1$)-electron systems. 

In addition to traditional $\Delta$ approaches,
many methods based on 
response theories have also been developed to calculate CLBEs. 
These methods, 
including configuration interaction methods\cite{asmurufCalculationNearedgeXray2008,
maganasRestrictedOpenShellConfiguration2014,
ehlertQuestBestSuited2017},
equation-of-motion coupled-cluster methods
\cite{nooijenDescriptionCoreExcitation1995,liuBenchmarkCalculationsKEdge2019},
and the Green's function methods\cite{barthManybodyTheoryCorevalence1981,
barthTheoreticalCorelevelExcitation1985,
ekstromPolarizationPropagatorXRay2006,
golzeAccurateAbsoluteRelative2020a,
zhu2021all,
liRenormalizedSinglesGreen2021,
li2022benchmark,
li2022renormalized,
li2023linear},
directly predict the charged excitation energies by 
calculating quasiparticle energies of the N-electron system.
Thus, the numerically challenging optimization of the core-ionized state 
can be avoided.
The $GW$ method \cite{hedinNewMethodCalculating1965}
is one of the most popular methods among these approaches.
By replacing the bare Coulomb interaction to the dynamical screened interaction,
$GW$ has become a promising method to predict CLBEs 
for both solid-state and molecular systems\cite{ishiiInitioGWQuasiparticle2001,
rostgaardFullySelfconsistentGW2010,
baumeierExcitedStatesDicyanovinylSubstituted2012,
carusoUnifiedDescriptionGround2012,
govoniLargeScaleGW2015,
maggioGWVertexCorrected2017,liRenormalizedSinglesGreen2021}.
Despite their great success, 
one-shot $GW$ approaches have strong XC functional dependence
\cite{fuchsQuasiparticleBandStructure2007},
making it difficult to find a XC functional that can provide accurate descriptions
for both core-level and valence-level electronic structures
\cite{dauthPiecewiseLinearityApproximation2016,
golzeAccurateAbsoluteRelative2020a}.
Self-consistent approaches, such as ev$GW$ and sc$GW$, can eliminate 
the XC functional dependence but significantly increase the computational cost
\cite{carusoSelfconsistentAllelectronImplementation2013,
carusoBenchmarkGWApproaches2016,kaplanQuasiParticleSelfConsistentGW2016}.

Approaches based on the density functional theory (DFT) have also been
designed to predict CLBEs. Chong et al.\cite{chongAccurateCalculationCoreelectron1995,
chongDensityFunctionalCalculation1995}
tested the performances of several generalized 
Slater's transition-state (GSTS) models\cite{slater1972advan,
williamsGeneralizationSlaterTransition1975} using B88/P86 functional
\cite{beckeDensityfunctionalExchangeenergyApproximation1988,
perdewDensityfunctionalApproximationCorrelation1986} and found that the 
unrestricted GSTS model provides the most accurate results. 
The statistical average of orbital potentials 
\cite{chongInterpretationKohnSham2002,
takahataDFTCalculationCoreelectron2003,
takahataAccurateCalculationN1s2008} was also applied to calculate 
CLBEs and achieved high accuracy. 
Crucially, 
the localization of core-level holes has been identified as a significant factor 
in calculating CLBEs.\cite{chongLocalizedDelocalized1s2007}
More recently, 
a new method employing generalized Kohn-Sham (GKS) orbital energies 
derived from the random phase approximation energy functional 
with a semicanonical projection (d-GKS-spRPA) was developed\cite{
vooraEffectiveOneparticleEnergies2019}.
This approach allows for the accurate determination of CLBEs from 
effective one-particle energies.


In this letter, we develop a new method for the accurate prediction of CLBEs 
from ground-state DFT calculations by considering both 
the localization of the core-level orbitals and the screening effect,
within the framework of the localized orbital scaling correction (LOSC)\cite{
liLocalizedOrbitalScaling2018,
suPreservingSymmetryDegeneracy2020,meiExactSecondOrderCorrections2021}.

The localization procedure 
and the curvature matrix capturing the screening interaction
were both developed in the scaling correction (SC) methods\cite{
liLocalizedOrbitalScaling2018,
suPreservingSymmetryDegeneracy2020,meiExactSecondOrderCorrections2021} 
designed to eliminate the delocalization 
error (DE)\cite{cohenFractionalSpinsStatic2008,
cohen2008insights,
mori-sanchezLocalizationDelocalizationErrors2008,
cohenChallengesDensityFunctional2012}
in commonly used DFAs. 
For small or moderate systems with fractional charges, 
conventional DFAs violate the Perdew-Parr-Levy-Balduz (PPLB) 
linearity condition\cite{perdewDensityFunctionalTheoryFractional1982,
yangDegenerateGroundStates2000}, 
which states that the ground-state energy 
should be piecewise linear between integer points. 
Conventional DFAs typically produce convex curves, 
underestimating ground-state energies for fractional charge systems. 
For large systems, 
the convex deviation decreases, 
and the DE exists in another way, 
which results in too low 
relative ground-state energies for ionized systems. 
To systematically reduce the DE,
researchers at the Yang laboratory developed various SC methods. 
The global scaling correction (GSC)\cite{zhengImprovingBandGap2011a} 
can restore the PPLB condition globally
by correcting the convex curve for
systems with fractional charges while preserving the ground-state energies for
integer systems. 
However, 
GSC's accuracy diminishes for larger systems where ground-state energies 
for integer systems are underestimated.
To address this, 
the local scaling correction (LSC) 
method\cite{liLocalScalingCorrection2015} was developed,
applying the energy correction locally.
The LOSC methods\cite{liLocalizedOrbitalScaling2018,
suPreservingSymmetryDegeneracy2020,
meiSelfConsistentCalculationLocalized2020,
mahlerLocalizedOrbitalScaling2022,
mei2022libsc} combine GSC and LSC
to achieve systematic improvement for descriptions of both small and large systems.
More recently, 
the GSC2 method \cite{meiExactSecondOrderCorrections2021}
was developed to capture exact second-order corrections to DFAs 
based on canonical molecular orbitals (CMOs), 
providing further improvement in descriptions of electronic structures 
of chemical systems.
The GSC2 energy correction can be written as
\begin{equation}
    \Delta_{\text{GSC2}}(\{n_{p\sigma}\}) = 
    \frac{1}{2}
    \sum_{p\sigma}
    \frac{\partial ^2 E(\{n_{m\tau}\})}{\partial n_{p\sigma}^2}
    (n_{p\sigma} - n_{p\sigma}^2),
    \label{eq:gsc2correction}
\end{equation}
where $n_{p\sigma}$ is the occupation number for the $p$th orbital with 
spin $\sigma$, and $E(\{n_{m\tau}\})$ is the energy of the system as a
function of the set of occupation numbers $\{n_{m\tau}\}$. 
The second-order 
derivative can be evaluated using the Maxwell relationship and linear response 
theory\cite{meiExactSecondOrderCorrections2021}
\begin{equation}
    \frac{\partial ^2 E(\{n_{m\tau}\})}{\partial n_{p\sigma} \partial n_{q\sigma}} = 
    \langle \psi_{p\sigma}\psi_{p\sigma}^*|
    K^{\sigma\sigma} + \sum_{\tau\nu} K^{\sigma\tau} \chi^{\tau\nu} K^{\nu\sigma}
    |\psi_{q\sigma}\psi_{q\sigma}^*\rangle,
    \label{eq:gsc2kappa}
\end{equation}
where $K^{\sigma\tau}$ represents Hartree-exchange-correlation kernels,
and $\chi^{\tau\mu}$ is the generalized linear response function
\cite{yang2012analytical,peng2013fukui}.

In connection to the chemical hardness, 
$\frac{\partial ^2 E(\{n_{m\tau}\})}{\partial n_{p\sigma}^2}$ is called 
the orbital hardness.
To understand the physical meaning of the orbital hardness, we write the 
associated 4-point generalized dielectric function $\varepsilon^{\sigma\tau}$
as $(\varepsilon^{-1})^{\sigma\tau}(\mathbf{r_1},\mathbf{r_2},\mathbf{r_3}
,\mathbf{r_4}) = 
\frac{\delta H_s^{\sigma}(\mathbf{r_1},\mathbf{r_2})}
{\delta v^{\tau}(\mathbf{r_3},\mathbf{r_4})}$. Now the expression in 
equation~\ref{eq:gsc2kappa} leads to
\begin{equation}
    W^{\sigma\zeta} = 
    K^{\sigma\zeta} + \sum_{\tau\nu} K^{\sigma\tau} \chi^{\tau\nu} K^{\nu\zeta}
    = \sum_\nu (\varepsilon^{-1})^{\sigma\nu}K^{\nu\zeta}
    \label{eq:orb_hardness}
\end{equation}
equation~\ref{eq:orb_hardness} can be interpreted as the generalized screened 
interaction\cite{martin2016interacting,meiExactSecondOrderCorrections2021}.

In this letter, we combine the localization procedure from the LOSC and 
the screened interaction from the GSC2 to develop a new approach, 
linear-response LOSC (lrLOSC),
for accurately predicting CLBEs.
The localization procedure is taken from the second version of 
LOSC (LOSC2)\cite{suPreservingSymmetryDegeneracy2020},
which preserves the symmetry and the degeneracy of chemical systems.
CMOs are linearly combined to minimize the
cost function to ensure both physical and energy space localization:
\begin{equation}
    F^\sigma = (1-\gamma)F_r^\sigma + \gamma C F_e^\sigma,
    \label{eq:cost_tot}
\end{equation}
where
$\gamma$ is set to be 0.707, 
and $C$ is set to be 1000 (in atomic units). 
$F_r^\sigma$ is the physical space cost function for spin $\sigma$ 
taken from the Foster-Boys 
localization\cite{boys1960construction},  
    $F_r^\sigma = 
    \sum_{p\sigma}[\langle \phi_{p\sigma}|\mathbf{r^2}|\phi_{p\sigma}\rangle
    -\langle \phi_{p\sigma}|\mathbf{r}|\phi_{p\sigma}\rangle ^2]$,
$F_e^\sigma$ is the energy space cost function for spin $\sigma$, 
    $F_e^\sigma = 
    \sum_{p\sigma}[\langle \phi_{p\sigma}|\hat{h}^2|\phi_{p\sigma}\rangle
    -\langle \phi_{p\sigma}|\hat{h}|\phi_{p\sigma}\rangle ^2]$,
and $\phi_{p\sigma}$ are orbitalets.
Orbitalets are localized orbitals that are localized in multiple spaces
obtained from linear combinations of both the occupied and the virtual CMOs, 
$\phi_{p\sigma} = \sum_q U_{pq}^\sigma\psi_{q\sigma}$.
In practice, 
energy windows are often applied to select CMOs to reduce the computational cost.

The screening effect in lrLOSC is included by adopting
the curvature matrix generalized from equation~\ref{eq:gsc2kappa},
\begin{equation}
    \kappa_{pq}^{\sigma} = 
    \langle \phi_{p\sigma}\phi_{p\sigma}^*|
    K^{\sigma\sigma} + \sum_{\tau\nu} K^{\sigma\tau} \chi^{\tau\nu} K^{\nu\sigma}
    |\phi_{q\sigma}\phi_{q\sigma}^*\rangle.
    \label{eq:losc22kappa}
\end{equation}
Instead of CMOs, orbitalets are adopted in equation~\ref{eq:losc22kappa}.
equation~\ref{eq:orb_hardness} remains valid, 
ensuring that the screening effect is accounted for in lrLOSC. 
Consequently, equation~\ref{eq:cost_tot} and equation~\ref{eq:losc22kappa} lead to the following 
correction to the orbital energies:
\begin{equation}
    \Delta \epsilon_{p\sigma} = 
    \sum_i \kappa_{qq} \left(\frac{1}{2}-\lambda_{qq}\right)|U_{qp}^\sigma|^2
    - \sum_{q\neq s}\kappa_{qs}^\sigma\lambda_{qs}^\sigma 
    U_{qp}^\sigma U_{sp}^{\sigma *},
    \label{eq:losc22delta_orb}
\end{equation}
where $\lambda_{pq}^\sigma$ is the local occupation matrix.
The correction to the total energy is 
\begin{equation}
    \Delta_{\text{lrLOSC}} = 
    \frac{1}{2}\sum_{pq\sigma}
    \kappa_{pq}^\sigma\lambda_{pq}^\sigma(\delta_{pq} - \lambda_{pq}^\sigma)
\end{equation}

CLBEs are approximated as effective one-particle energies
following the methodology described 
in Refs.~\citenum{cohen2008fractional,meiApproximatingQuasiparticleExcitation2019,
vooraEffectiveOneparticleEnergies2019,li2022combining}.
Orbitals with energy differences of less than 0.05 eV are considered degenerate, 
and the quasihole energy $\omega^-(N)$ describing the removal of one core-level electron
from an $N$-electron system
is approximated as the average of 
the degenerate orbital energies $\{\varepsilon_i(N)\}$, i.e.
\begin{equation}
    \omega^-(N) = E_0(N) - E(N-1) 
                \approx \frac{1}{g} \sum_{i=1}^g \varepsilon_i(N),
\end{equation}
where $g$ is the number of degenerate core-level orbitals, 
$E_0(N)$ represents the ground-state energy of the $N$-electron system,
and $E(N-1)$ represents the energy of the ($N-1$)-electron system with
a core-level hole.

The MAEs and the mean signed errors (MSEs) of absolute and relative CLBEs 
for different core species of systems from CORE65 test set\cite{
golze2020accurate} are shown in
Table~\ref{tab:abs} and Table~\ref{tab:rel} respectively.
The first thing we noticed is that MAEs from lrLOSC are the lowest in both 
the tables for all the functionals, 
indicating the ability of lrLOSC to improve the accuracy of CLBE calculations.
The high accuracy of lrLOSC results is due to 
a) the proper localization of the core-level states
and 
b) the well described screening effect.

We begin with the discussion of the influence of the localization procedure.
As shown in Table~\ref{tab:rel}, common DFAs can provide fairly accurate 
relative CLBEs.
The MAEs from BLYP, B3LYP and PBE 
calculations are about 0.7 eV. 
However, common DFAs fail for absolute CLBEs,
as shown in Table~\ref{tab:abs},
all studied DFAs underestimate the absolute CLBEs by over 15 eV.
The fact that DFAs can provide much more accurate relative 
CLBEs than absolute CLBEs indicates that absolute CLBEs from DFA calculations 
need to be shifted. 
This shift can be achieved by correcting the DE.
As shown by the MAEs and MSEs in Table~\ref{tab:abs},
LOSC2 \cite{suPreservingSymmetryDegeneracy2020} can shift the calculated CLBEs by around 30 eV.
Overall, LOSC2 overestimate the absolute CLEBs by excessively 
shifting the energy levels.
This is similar to the LOSC2 over correction for valence orbital energies 
for bulk systems \cite{mahlerLocalizedOrbitalScaling2022}.

\begin{figure}
    \centering
    \subfigure[errors of absolute CLBEs]
    {\includegraphics[width=0.8\linewidth]{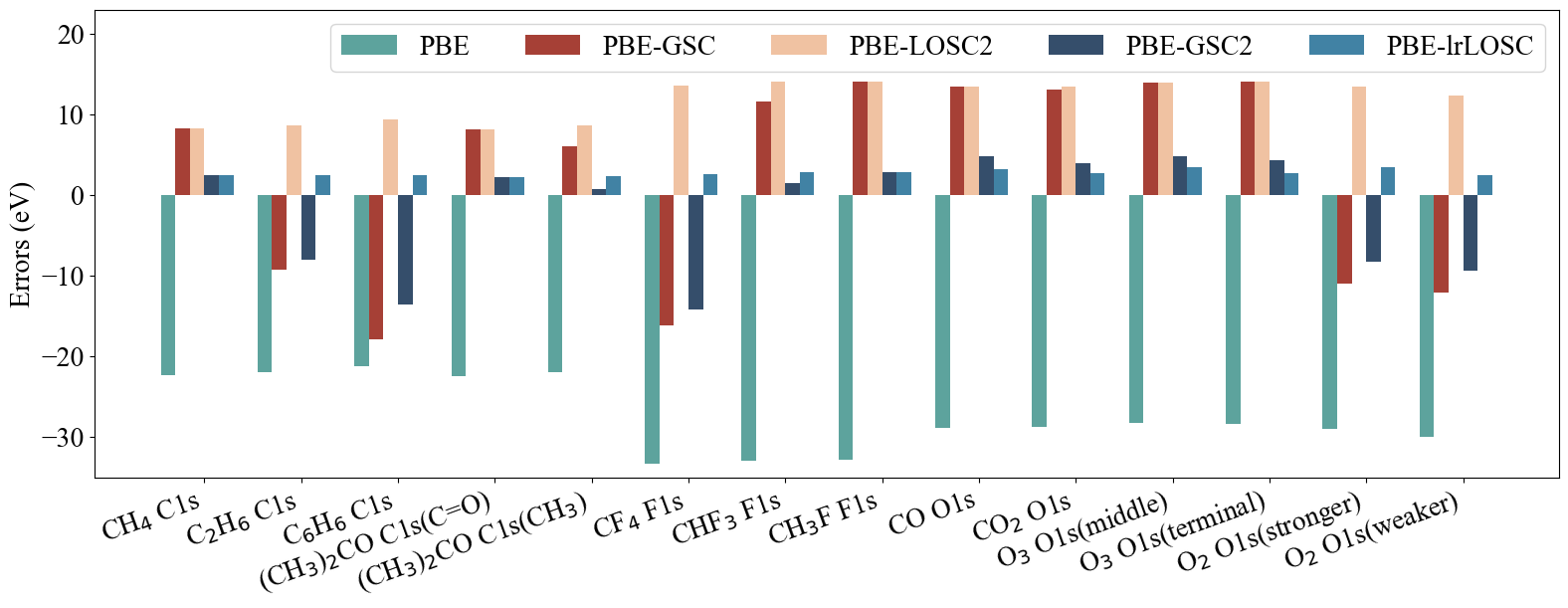}}\\
    \subfigure[errors of relative CLBEs]
    {\includegraphics[width=0.8\linewidth]{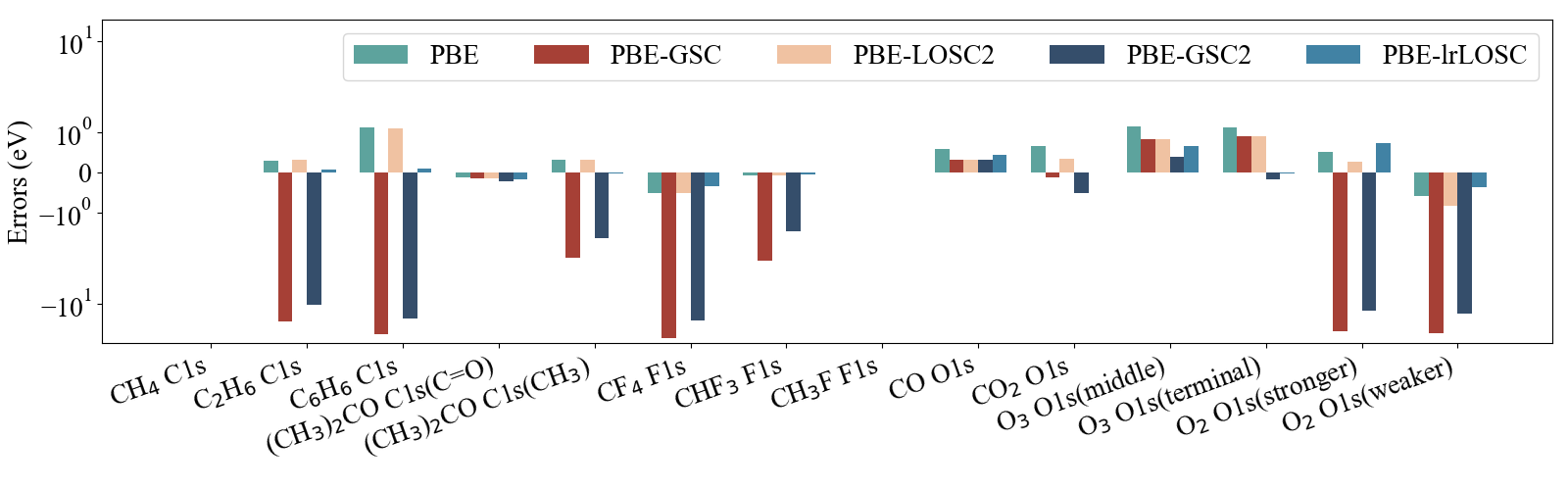}}
    \caption{
        Errors of (a) absolute and (b) relative CLBEs of systems from 
        CORE65 test set. The calculations were done 
        with PBE/def2-TZVP.
        The vertical axis of subfigure (b) uses
        symmetrical logarithmic scale.
    }
    \label{fig:error}
\end{figure}

On the other hand,
the localization procedure also plays an important role in the prediction 
of relative CLBEs.
As tabulated in Table~\ref{tab:rel},
the MAEs from LOSC2 calculations can be around 3 eV lower than 
those from GSC \cite{zhengImprovingBandGap2011a} calculations, 
and the MAEs from lrLOSC calculations are over 90\% lower than 
those from GSC2 \cite{meiExactSecondOrderCorrections2021} calculations.
The localization is especially important for systems with degenerate core-level states.
To further illustrate this,
the errors of absolute and relative CLBEs from PBE calculations for 14 
selected systems are plotted in Figure~\ref{fig:error}.
Among these systems, \ce{C2H6}, \ce{C6H6}, \ce{(CH3)2CO}, \ce{CF4}, \ce{CHF3},
\ce{CO2} and \ce{O3}
have degenerate core-level states due to structural symmetry,
while the degenerate core-level states of \ce{O2} arises from the parent DFA 
calculation.
For these systems,
the localization of core-level holes is important for the calculation of 
CLBEs. 
However, the degenerate core-level orbitals from conventional DFT calculations 
can often be delocalized.
Adding GSC or GSC2 corrections based on these delocalized orbitals leads to 
inaccurate results.
We can take \ce{C2H6} that has two degenerate core-level C1s states as 
an example.
As shown in Figure~\ref{fig:error} (b), 
the absolute errors from the GSC and the GSC2 calculations are significantly larger 
than those from the LOSC2 and the lrLOSC calculations.
On the other hand,
\ce{CH4}, 
which shares similar size and chemical properties with \ce{C2H6} 
but lacks degenerate core-level orbitals,
has weaker dependency on the localization procedure.
As shown in Figure~\ref{fig:error} (a), for \ce{CH4},
the errors from GSC and LOSC2 are similar,
and the errors from GSC2 and lrLOSC are also very close.
For \ce{C6H6}, 
with an even greater number of degenerate states, 
the difference between GSC2 and lrLOSC results is more significant.
For systems such as \ce{CO2} and \ce{CHF3},
although degenerate core-level states exist, 
the corresponding atoms are not directly connected with a chemical bond.
Therefore, 
the influence of the localization on these systems is less significant
but still exists.
Figure~\ref{fig:error} (b) demonstrates that the delocalization nature 
of CMOs poses more significant issues in calculating relative CLBEs. 
GSC and GSC2 results for systems with degenerate core-level states exhibit 
substantial deviations from the reference values.

Describing the screening effect is also very important for predicting CLBEs.
In GSC2 and lrLOSC, 
the screening effect is well captured by the curvature matrices 
as in equation~\ref{eq:gsc2kappa} and equation~\ref{eq:losc22kappa}
respectively.
As shown in Table~\ref{tab:abs},
both MAE and MSEs from GSC2 and lrLOSC are significantly lower than those 
from GSC and LOSC2.
As a result,
the overcorrection introduced by the localization procedure mentioned earlier 
can be significantly alleviated in lrLOSC.
For each functional, the MAEs from lrLOSC are reduced to about one third 
of those from LOSC2.
The importance of incorporating the screening effect is also shown by
the plotted errors in Figrue~\ref{fig:error} (a),
where the absolute heights of GSC2 and lrLOSC bars are much lower than those 
of GSC and LOSC2 bars.
By effectively capturing the screening effect, 
lrLOSC also provides more accurate relative CLBEs. 
For example, the MAE of B3LYP-lrLOSC is 0.14 eV, 
which is 0.46 eV lower than the MAE of B3LYP-LOSC2 results.

The screening effect is significantly more important for core-level orbitals 
compared to valence orbitals. 
This difference can be explained by 
considering the size and environment of the orbitals. 
Because the core-level orbitals
are localized at the center of molecules and entirely surrounded by other orbitals, 
the removal of a core-level electron alters the environment for all other orbitals, 
which then relax or adjust to the formation of the core hole.
This is the screening effect, 
which can be effectively captured with lrLOSC. 
In contrast, 
the effect is less pronounced for a valence hole, 
as valence shells are much larger and surrounded by only the vacuum outside molecules.

In summary, a new method, lrLOSC, 
has been developed for predicting CLBEs based on ground-state DFT calculations. 
By incorporating both the localization of core-level holes and the screening effect, 
lrLOSC significantly improves the accuracy of CLBE predictions 
compared to parent DFA calculations. 
The comparison of lrLOSC with LOSC2 and GSC2 results indicates that 
both the localization of core-level holes and the accurate description of 
screening effect are crucial for precise CLBE calculations. 
The localization procedure can introduce a positive energy shift on the basis
of DFA calculations and is essential for predicting accurate relative CLBEs 
especially for systems with degenerate core-level states.
Capturing the screening effect by incorporating the generalized GSC2 curvature 
matrix universally improves the prediction of both absolute and relative CLBEs.
Building on the success of lrLOSC in calculating CLBEs, 
we plan to further investigate the effectiveness of lrLOSC 
in providing accurate valence and near-valence orbital energies, 
quasiparticle energies, and photoemission spectra in future studies.

\begin{table}
\setlength\tabcolsep{2pt}
\footnotesize
    \centering
    \begin{threeparttable}
        \caption{MAEs of absolute CLBEs for different 
        core atoms (in eV) of systems from CORE65 test
        set\cite{golze2020accurate}. The calculations were performed using def2-TZVP basis 
        set\cite{weigend2005a} with QM4D\cite{qm4d}.}
        \label{tab:abs}
        \begin{tabular}{c|rrrrrr|rrrrrr|rrrrrr}
             \cline{1-19}
             \multirow{2}{*}{species} & \multicolumn{6}{c|}{BLYP} & \multicolumn{6}{c|}{B3LYP} & \multicolumn{6}{c}{PBE}\\
             \cline{2-7}\cline{7-13}\cline{12-19}
                            &   A  &  B   &    C   &  D     &    E   &F     &   A  &  B   &    C   &   D   &   E     &F     &   A  &  B   &    C  &  D     & E     & F\\
             \cline{1-19}                                                                                                                                  
             C              &21.86 &8.88  & 8.74   & 3.74   & 2.87   & 2.98 &14.48 & 8.87 & 10.02  & 3.26  & 2.86    & 2.98 & 22.56& 8.30 & 8.04  & 3.34   & 2.34  & 2.46\\
             F              &32.27 &14.60 & 14.62  & 3.46   & 3.47   & 4.18 &21.32 &9.14  & 16.20  & 4.86  & 2.78    & 3.49 &33.00 & 13.94& 13.89 & 6.15   & 2.76  & 3.47\\
             N              &24.50 & 11.59& 11.54  & 3.65   & 3.15   & 3.39 &15.95 &11.34 & 12.89  & 3.20  & 2.84    & 3.07 &25.07 & 11.13& 10.97 & 3.30   & 2.69  & 2.93\\
             O              &27.83 &13.99 & 14.65  & 5.42   & 3.42   & 3.84 &18.03 &13.75 & 16.12  & 4.33  & 2.92    & 3.35 &28.45 & 13.80& 13.95 & 4.85   & 2.86  & 3.29\\
             \cline{1-19}                                                                                                                                
             MAE\tnote{a}\ \ &24.97   &11.45 & 11.61  & 4.30   & 3.14   & 3.42 &16.34    &11.03 & 12.99  & 3.71  & 2.88    & 3.15 & 25.62  & 11.02& 10.94 & 4.01   & 2.61  & 2.89\\
             MSE\tnote{a}\ \ &$-$24.97& 8.58 & 11.61  & 2.04   & 3.14   & 3.42 &$-$16.34 & 8.96 & 12.99  & 1.60  & 2.88    & 3.15 &$-$25.62&  7.96& 10.94 & 1.32   & 2.61  & 2.89\\
             \cline{1-19}
        \end{tabular}
        \begin{tablenotes}
        \item[a] The MAEs and the MSEs are calculated from CORE65 test set\cite{golze2020accurate}. Nitrobenzene and 
        phenylacetylene data are not calculated considering computational cost.\\
        \item[b] A = KS-DFT, B = GSC, C = LOSC2, D = GSC2, E = lrLOSC, F = lrLOSC
        with relativistic correction\cite{keller2020relativistic}.
        \end{tablenotes}
    \end{threeparttable}
\end{table}

\begin{table}
\setlength\tabcolsep{4pt}
\footnotesize
    \centering
    \begin{threeparttable}
        \caption{MAEs of relative CLBEs
        for different core atoms
        (in eV) of systems from CORE65 test
        set\cite{golze2020accurate}. The calculations were performed using def2-TZVP basis 
        set\cite{weigend2005a} with QM4D\cite{qm4d}. The relative CLBE
        is defined as the energy difference relative to a reference molecule, 
        $\Delta E_{\text{relative}}=E_{\text{absolute}} - E_{\text{ref\_absolute}}$. 
        \ce{CH4}, \ce{NH3}, \ce{H2O} and \ce{CH3F} are the reference molecules for 
        C1s, N1s, O1s and F1s respectively. }
        \label{tab:rel}
        \begin{tabular}{c|rrrrr|rrrrr|rrrrr}
             \cline{1-16}
             \multirow{2}{*}{species} & \multicolumn{5}{c|}{BLYP} & \multicolumn{5}{c|}{B3LYP} & \multicolumn{5}{c}{PBE}\\
             \cline{2-6}\cline{7-11}\cline{12-16}
                            & A    & B    &    C   &  D     & E      &   A  &  B   &    C   &  D    & E       &   A  & B    &    C  & D      & E      \\
             \cline{1-16}                                                                                     
             C              &0.52  &3.68  & 0.52   & 2.06   & 0.14   &0.41  & 3.49 & 0.41   & 1.67  & 0.11    & 0.55 & 3.78 & 0.56  & 2.13   & 0.16\\
             F              &0.17  &0.19  & 0.17   & 0.11   & 0.10   &0.12  &12.78 & 0.12   & 5.78  & 0.07    &0.19  & 10.94& 0.19  & 6.16   & 0.13\\
             N              &0.62  & 2,68 & 0.62   & 1.22   & 0.10   &0.64  &3.88  & 0.64   & 1.56  & 0.05    &0.69  & 2.73 & 0.69  & 1.21   & 0.06\\
             O              &0.99  &4.24  & 0.80   & 2.06   & 0.20   &1.03  &4.58  & 0.86   & 1.79  & 0.22    &1.04  & 3.27 & 0.86  & 1.51   & 0.22\\
             \cline{1-16}
             MAE\tnote{a}\ \ &0.68 &3.52     & 0.62   & 1.81     & 0.15     &0.65  &4.43     & 0.60   & 1.91    & 0.14    & 0.73 & 3.79   & 0.67  & 1.96     & 0.16\\
             MAE\tnote{a}\ \ &0.28 &$-$2.83  & 0.21   &$-$1.78   &$-$0.06   &0.39  &$-$3.71  & 0.32   &$-$1.80  & 0.07    & 0.31 &$-$3.01 & 0.23  &$-$1.89   &$-$0.01\\
             \cline{1-16}
        \end{tabular}
 
        \begin{tablenotes}
        \item[a] The MAEs and the MSEs are calculated from CORE65 test set\cite{golze2020accurate}. Nitrobenzene and 
        phenylacetylene data are not calculated considering computational cost.\\
        \item[b] A = KS-DFT, B = GSC, C = LOSC2, D = GSC2, E = lrLOSC.
        \end{tablenotes}
    \end{threeparttable}
\end{table}

\FloatBarrier

\begin{acknowledgement}
The authors acknowledge the support from 
the National Science Foundation (Grant No. CHE-2154831).
\end{acknowledgement}

\begin{suppinfo}
Detailed absolute and relative CLBEs from GSC, LOSC2, GSC2, and lrLOSC using
different DFAs.
\end{suppinfo}

\bibliography{CoreBindingEnergy-Theo,CoreBindingEnergy-Exp,DFA,SC,extra}

\end{document}